\documentclass[10pt,twocolumn,amsmath,amssymb,aps,prl,showpacs,longbibliography,superscriptaddress]{revtex4-1}
\usepackage[latin9]{inputenc}
\usepackage{textcomp}
\usepackage{amsmath}
\usepackage{amssymb}
\usepackage{graphicx}
\usepackage{esint}

\makeatletter

\newcommand*\LyXThinSpace{\,\hspace{0pt}}

\usepackage{amsfonts}
\usepackage{graphics}

\makeatother

\begin{document}
\title{Quantum Lifshitz points in an altermagnetic metal}
\author{Hui Hu}
\affiliation{Centre for Quantum Technology Theory, Swinburne University of Technology,
Melbourne 3122, Australia}
\author{Xia-Ji Liu}
\affiliation{Centre for Quantum Technology Theory, Swinburne University of Technology,
Melbourne 3122, Australia}
\date{\today}
\begin{abstract}
We predict the existence of two tri-critical quantum Lifshitz points
in recently discovered $d$-wave altermagnetic metals subjected to
an external magnetic field. These points connect a spatially modulated
Fulde--Ferrell--Larkin--Ovchinnikov (FFLO) phase, a uniform polarized
Bardeen--Cooper--Schrieffer (BCS) superconducting phase, and the
normal metallic phase in a nontrivial manner. Depending on whether
the FFLO state is primarily induced by the magnetic field or by $d$-wave
altermagnetism, we classify the corresponding Lifshitz points as field-driven
or altermagnetism-driven, respectively. Notably, the two types exhibit
distinct behaviors: the transition from the FFLO phase to the polarized
BCS phase is first-order near the field-driven Lifshitz point, as
might be expected, whereas it becomes continuous near the altermagnetism-driven
Lifshitz point. We further explore the effects of finite temperature
and find that the altermagnetism-driven Lifshitz point is significantly
more sensitive to thermal fluctuations.
\end{abstract}
\maketitle
Understanding phase transitions between different competing quantum
orders lies at the heart of diverse fields \citep{Sachdev1999,Sachdev2023},
including quantum many-body physics, condensed matter physics, and
high-energy particle physics. In this regard, a special critical point
known as the tri-critical Lifshitz point attracts specific attention
over the last several decades, due to its unusual critical exponents
\citep{Hornreich1975,Sak1978,Mergulhao1999,Diehl2000,Pisarski2019}.
The Lifshitz point arises when three phases meet together: two ordered
phases (one uniform and another unevenly distributed in real space)
coexisting with a disordered phase. Typically, Lifshitz points are
classical and present at nonzero temperature ($T>0$), where the disordered
to ordered transitions are often driven by thermal fluctuations \citep{Hornreich1975}.
The most known examples include helimagnetic materials such as MnP
nanorod films \citep{Madhogaria2021}, inhomogeneous polymers \citep{Fredrickson1997,Jones2012},
ferroelectric chiral smectic liquid crystals \citep{Musevic1982},
spin-population imbalanced Fermi gases \citep{Gubbels2009}, and even
hot dense quantum chromodynamics (QCD) matters \citep{Pisarski2019,Casalbuoni2004}.

In principle, the Lifshitz point can also emerge at zero temperature
($T=0$) \citep{Ramazashvili1999}, where phase transitions are governed
solely by quantum fluctuations \citep{Sachdev1999}. Nonetheless,
identifying such a \emph{quantum }Lifshitz point in real materials
has proven to be exceptionally challenging. To date, theoretical studies
of quantum Lifshitz points have primarily focused on some idealized
models or systems, such as the infinite-dimensional Hubbard model
\citep{Gunther2007}, the frustrated ferromagnetic Heisenberg chain
\citep{Balents2016}, the two-dimensional (2D) effective theory with
specifically engineered interaction potential \citep{Wardh2018},
the spin-1/2 square-lattice $J$-$Q$ model \citep{Zhao2020}, the
frustrated 2D XY model \citep{Kharkov2020}, and Fermi mixtures with
both spin-imbalance and mass-imbalance \citep{Zdybel2020}. 

In this Letter, we would like to propose that, the recently discovered
\emph{altermagnetic} metals may provide an ideal avenue to realize
and observe the long-sought quantum Lifshitz point and to explore
its unconventional critical exponents. Altermagnetism represents a
distinct form of magnetic order, different from the well-known ferromagnetism
and antiferromagnetism \citep{Hayami2019,Hayami2020a,Hayami2020b,Smejkal2020,Smejkal2022,Mazin2023,Das2024,Hiraishi2024,Amin2024}.
In altermagnetic metals, collinear spin arrangements result in zero
net magnetization; however, due to specific crystal symmetries, significant
spin splitting still appears in the electronic band structure. This
unique spin-splitting behavior can have profound effects when inter-particle
interactions are considered \citep{McClarty2024,Lu2024,Maeda2024,Fukaya2025a,Fukaya2025b,Lin2025}.
Of particular relevance to our study is the recent surprising discovery
\citep{Chakraborty2024,Hong2025} of an altermagnetism-induced spatially
inhomogeneous Fulde--Ferrell--Larkin--Ovchinnikov (FFLO) phase
\citep{Fulde1964,Larkin1964} under attractive interactions.

Here, we compare the FFLO state induced by altermagnetism \citep{Chakraborty2024,Hong2025}
with the conventional FFLO state driven by an external magnetic field
\citep{Fulde1964,Larkin1964,Ohashi2002,Uji2006,Hu2006,Hu2007,Sheehy2015}.
Surprisingly, we discover that altermagnetism and the magnetic field
actually compete with one another in promoting the spatially modulated
superconducting FFLO phase. This competition gives rise to a homogeneous,
spin-population-polarized Bardeen--Cooper--Schrieffer (BCS) phase,
emerging from their combined influence. This newly identified polarized
BCS phase represents a uniform ordered state, whereas the two FFLO
phases --- one arising from altermagnetism and the other from a magnetic
field --- constitute exotic ordered states characterized by broken
spatial uniformity. Furthermore, recognizing that a disordered normal
state inevitably appears under sufficiently strong altermagnetism
or magnetic field, even at zero temperature, we identify the presence
of two quantum Lifshitz points. Remarkably, these arise naturally
without requiring fine-tuning of system parameters such as the mass
imbalance or the specific form of the interaction potential.

The emergence of quantum Lifshitz points in altermagnetic metals offers
a rare opportunity to deepen our understanding of the unusual universal
critical exponents characteristic of the Lifshitz regime \citep{Hornreich1975,Sak1978,Mergulhao1999,Diehl2000},
where quantum fluctuations are expected to play a pivotal role \citep{Pisarski2019}.
At the mean-field level, we already distinguish the distinct nature
of the two predicted quantum Lifshitz points, as the phase transition
between the two ordered superconducting states exhibits different
behaviors --- one being a discontinuous first-order transition and
the other a smooth second-order transition. We hypothesize that these
quantum Lifshitz points may remain robust in the presence of quantum
fluctuations, a question that calls for more comprehensive theoretical
exploration. We also anticipate that future experiments will uncover
the intriguing and anomalous quantum critical phenomena associated
with these transitions.

\textit{Model.} -- We consider a 2D spin-1/2 interacting Fermi system
with $d$-wave altermagnetism in a unit area subjected to an external
magnetic field $h$, as described by a Hamiltonian density ($s_{\uparrow}=+1$
and $s_{\downarrow}=-1$),

\begin{equation}
\mathcal{H}=\sum_{\sigma=\uparrow,\downarrow}\psi_{\sigma}^{\dagger}\left[\hat{\xi}_{\mathbf{k}}+s_{\sigma}\left(\hat{J}_{\mathbf{k}}+h\right)\right]\psi_{\sigma}^{\dagger}+U_{0}\psi_{\uparrow}^{\dagger}\psi_{\downarrow}^{\dagger}\psi_{\downarrow}\psi_{\uparrow},
\end{equation}
where $\psi_{\sigma}(\mathbf{x})$ is the annihilation field operator
for a fermion in the spin state $\sigma$, $\hat{J}_{\mathbf{k}}=\lambda\hbar^{2}\hat{k}_{x}\hat{k}_{y}/(2m)$
with $\hat{k}_{x,y}\equiv-i\partial_{x,y}$ and the dimensionless
coupling constant $\lambda$ is the spin-splitting due to the $d$-wave
altermagnetism \citep{Smejkal2022,Hong2025} and $\hat{\xi}_{\mathbf{k}}=-\hbar^{2}\nabla^{2}/(2m)-\mu$
is the conventional dispersion relation, with chemical potential $\mu$
that fixes the average density $n$. For simplicity, we consider an
$s$-wave attractive contact interaction, with its strength $U_{0}=-\sum_{\mathbf{k}}1/[\hbar^{2}k^{2}/m+\varepsilon_{B}]<0$
characterized by a two-body binding energy $\varepsilon_{B}$ \citep{He2015}.
Hereafter, the wavevector $k$ and energy (such as $h$ and $\varepsilon_{B}$)
are measured in units of the Fermi wavevector $k_{F}=(2\pi n)^{1/2}$
and Fermi energy $\varepsilon_{F}=\hbar^{2}k_{F}^{2}/(2m)=k_{B}T_{F}$,
respectively. Throughout the paper, we take a binding energy $\varepsilon_{B}=0.08\varepsilon_{F}$.

\textit{Pairing instability}. -- At sufficiently low temperature,
the attractive interaction drives the spin-1/2 Fermi system into a
superconducting state. This pairing instability is mostly easily seen
by introducing a pairing field $\Delta(\mathbf{x},\tau)$ through
the standard Hubbbard-Stratonovich transformation, which couples to
$\psi_{\uparrow}^{\dagger}\psi_{\downarrow}^{\dagger}$ \citep{He2015,SadeMelo1993}.
By integrating out the fermionic field operators and truncating to
fourth order in the pairing field $\Delta(Q\equiv\{\mathbf{q},i\omega_{m}\})$
in momentum space, where $\omega_{m}\equiv2m\pi k_{B}T$ with integer
$m=0,\pm1,\cdots$ is the bosonic Matsubara frequency, we obtain \citep{SadeMelo1993},
\begin{equation}
\mathcal{S}_{\textrm{eff}}=\sum_{Q}\frac{\bar{\Delta}_{Q}\Delta_{Q}}{-\Gamma\left(Q\right)}+\sum_{Q_{1}Q_{2}Q_{3}}\frac{u_{1,2,3}}{2}\Delta_{1}\bar{\Delta}_{2}\Delta_{3}\bar{\Delta}_{1+2-3}.\label{eq:Seff}
\end{equation}
Here, $\bar{\Delta}_{1+2-3}$ is an abbreviation of $\bar{\Delta}(Q_{1}+Q_{2}-Q_{3})$
and the sum $\sum_{Q}$ stands for $k_{B}T\sum_{m}\int d\mathbf{q}/(2\pi)^{2}$.
The inverse vertex function $\Gamma^{-1}(Q)=U_{0}^{-1}+\chi_{\textrm{pair}}^{0}(Q)$
can be calculated by summing all the ladder diagrams for the pair
propagator $\chi_{\textrm{pair}}^{0}(Q)$ \citep{SadeMelo1993},
\begin{equation}
\Gamma^{-1}\left(Q\right)=\frac{1}{U_{0}}+\sum_{\mathbf{k}}\frac{f\left(\xi_{\mathbf{q}/2+\mathbf{k}\uparrow}\right)+f\left(\xi_{\mathbf{q}/2-\mathbf{k}\downarrow}\right)-1}{i\omega_{m}-\left(\xi_{\mathbf{q}/2+\mathbf{k}\uparrow}+\xi_{\mathbf{q}/2-\mathbf{k}\downarrow}\right)},
\end{equation}
where $\xi_{\mathbf{k}\uparrow}=\xi_{\mathbf{k}}+J_{\mathbf{k}}+h$
and $\xi_{\mathbf{k}\downarrow}=\xi_{\mathbf{k}}-J_{\mathbf{k}}-h$,
and $f(E)\equiv(e^{E/k_{B}T}+1)^{-1}$ is the Fermi-Dirac distribution
function. The quartic term in Eq. (\ref{eq:Seff}) represents an effective
repulsion between Cooper pairs at the lowest order \citep{He2015,SadeMelo1993}.
In this Born approximation, it suffices to suppress the momentum dependence
of the coefficient $u_{1,2,3}$ \citep{He2015,SadeMelo1993}: $u\equiv u(Q_{1}=0,Q_{2}=0,Q_{3}=0)=\sum_{K}G_{0\uparrow}^{2}(K)G_{0\downarrow}^{2}(-K)>0$,
where $G_{0\sigma}(K)$ is the non-interacting Green function of fermions
at the four-momentum $K\equiv\{\mathbf{k},i\omega_{n}\}$.

\begin{figure}
\begin{centering}
\includegraphics[width=0.5\textwidth]{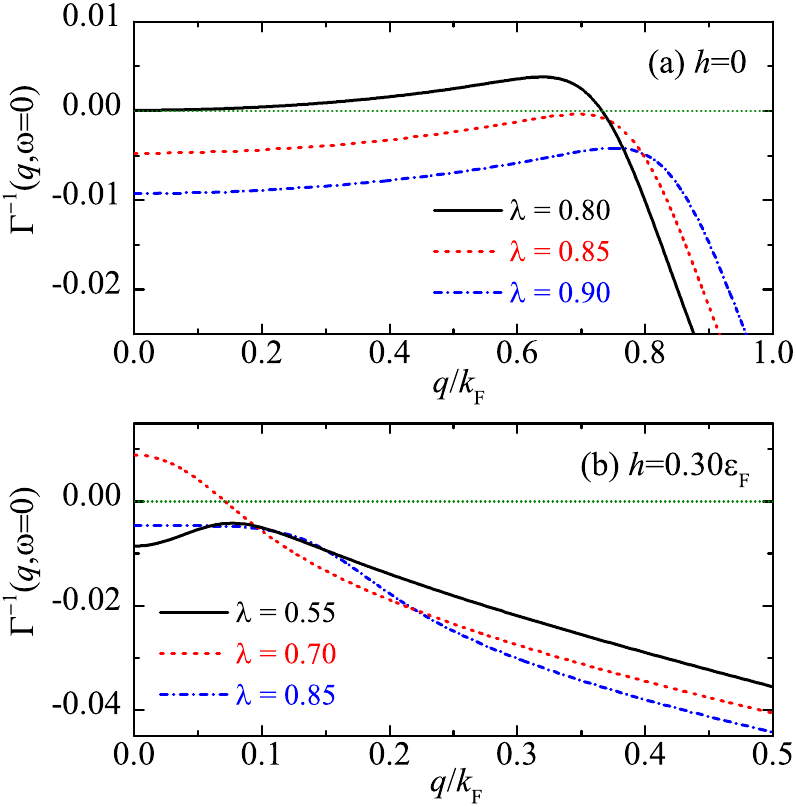}
\par\end{centering}
\caption{\label{fig1_instability} The pairing momentum $q$-dependence of
the inverse vertex function at zero frequency $\Gamma^{-1}(q,\omega=0)$,
in units of $2m/\hbar^{2}$, at different altermagnetic couplings
$\lambda$ as indicated, and at the magnetic field $h=0$ (a) and
$h=0.3\varepsilon_{F}$ (b). We have taken a negligible temperature
$T=0.01T_{F}$ to smooth the sharp Fermi surface and to increase numerical
accuracy.}
\end{figure}

\begin{figure*}
\begin{centering}
\includegraphics[width=0.3\textwidth]{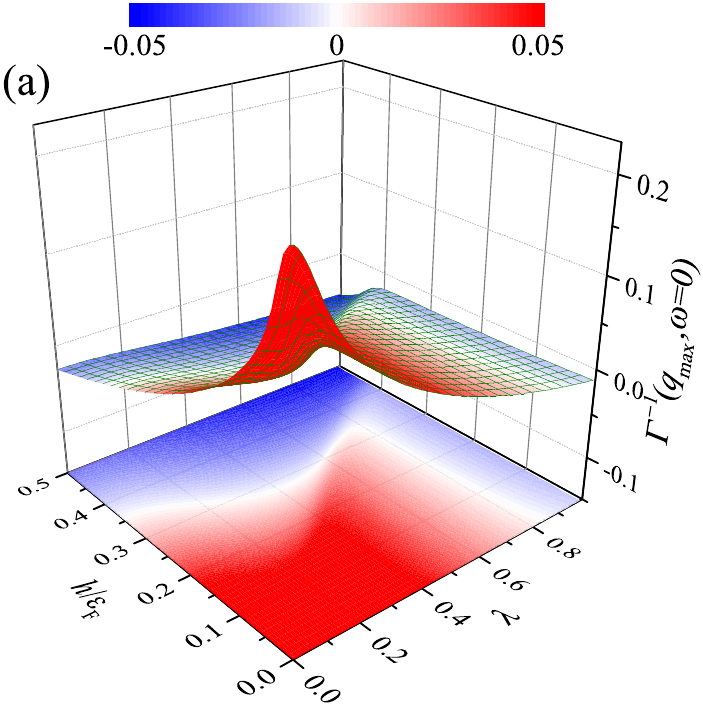}\includegraphics[width=0.3\textwidth]{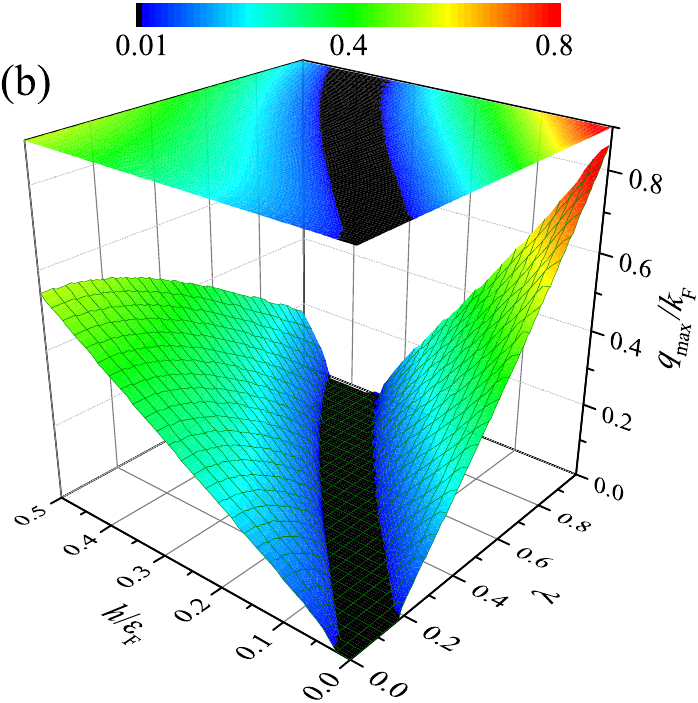}\includegraphics[width=0.4\textwidth]{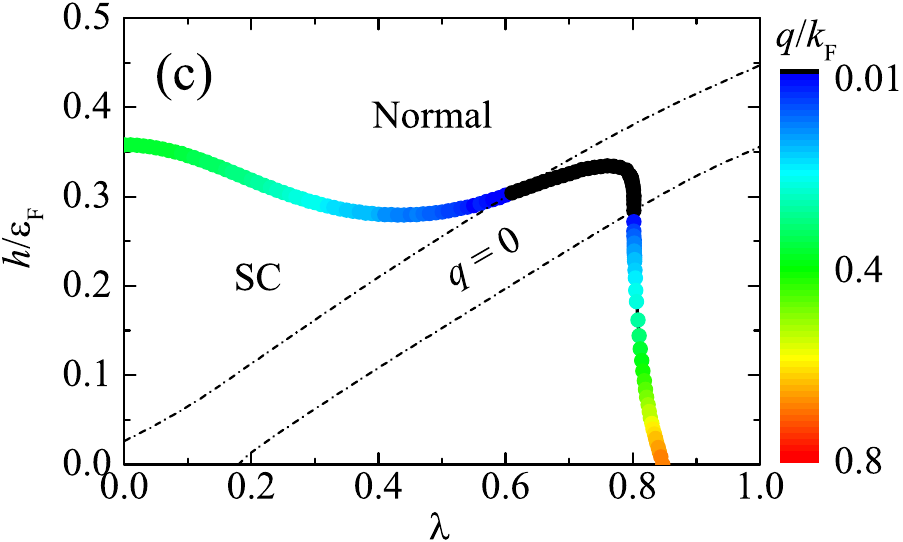}
\par\end{centering}
\caption{\label{fig2_invg0_qmax} (a) The maximum of the inverse vertex function
$\Gamma^{-1}(q_{\max},\omega=0)$, as functions of the altermagnetic
coupling $\lambda$ and magnetic field $h/\varepsilon_{F}$. (b) The
corresponding pairing momentum $q_{\max}$. (c) The instability line
determined by the thouless criterion $\Gamma^{-1}=0$ separates the
superconducting phase (SC) from the normal state. The color of the
line shows the pairing momentum. The pairing momentum is identically
zero between the two dot-dashed lines.}
\end{figure*}

The superconducting instability is determined by the well-known thouless
criterion ($i\omega_{m}\rightarrow\omega+i0^{+}$), 
\begin{equation}
\max_{\{\mathbf{q}\}}\Gamma^{-1}(\mathbf{q},\omega=0)\left|_{T=T_{c}}\right.=0,
\end{equation}
generalized here to allow a nonzero center-of-mass momentum $\mathbf{q}\neq0$
of Cooper pairs, which signals a spatially inhomogeneous FFLO superconducting
state \citep{Liu2006}. In Fig. \ref{fig1_instability}, we report
the $q$-dependence of the inverse vertex function $\Gamma^{-1}(q=\mathbf{\left|q\right|},\omega=0)$,
with varying altermagnetic coupling $\lambda$ at the two magnetic
fields, $h=0$ (a) and $h=0.3\varepsilon_{F}$ (b), and at essentially
zero temperature $T=0.01T_{F}$. At zero magnetic field in Fig. \ref{fig1_instability}(a),
the maximum of $\Gamma^{-1}(q,\omega=0)$ always occurs at a finite
momentum $q_{\max}\neq0$ and becomes non-negative once $\lambda\geq\lambda_{c}\simeq0.85$,
indicating the onset of an altermagnetism-induced FFLO state. This
result aligns with previous mean-field calculations \citep{Chakraborty2024,Hong2025},
despite differences in the model Hamiltonians and interaction types
employed. In stark contrast, at a finite magnetic field $h=0.3\varepsilon_{F}$
as shown in Fig. \ref{fig1_instability}(b), the position of the maximum
$q_{\max}$ exhibits a non-monotonic dependence on increasing $\lambda$.
While the inverse vertex function reaches its maximum at a nonzero
momentum $q_{\max}$ for both relatively weak and strong altermagnetic
couplings, it peaks precisely at zero momentum when $\lambda=0.70$.
This suggests that, at this intermediate coupling constant, the system
favors a BCS-like superconducting state that is spatially uniform.

\textit{Quantum Lifshitz point}s. -- To understand this non-monotonic
behavior, it is instructive to Taylor-expand the inverse vertex function,
keeping the leading orders in powers of $q$ and $\omega$, i.e.,
$-\Gamma^{-1}(q,\omega)=r+Zq^{2}+Dq^{4}-\gamma\omega$, where the
expansion coefficients $\{Z,D>0,\gamma\}$ as well as $r=-\Gamma^{-1}(0,0)$
are functions of the altermagnetic coupling $\lambda$ and magnetic
field $h$. Therefore, we may directly write down a Ginzburg-Landau
(GL) Lagrangian at the low-energy from $\mathcal{S}_{\textrm{eff}}$
in Eq. (\ref{eq:Seff}),
\begin{equation}
\mathcal{L}=\gamma\bar{\Delta}\partial_{\tau}\Delta+Z\left|\partial_{\mathbf{x}}\Delta\right|^{2}+D\left|\partial_{\mathbf{x}}^{2}\Delta\right|^{2}+r\left|\Delta\right|^{2}+\frac{u}{2}\left|\Delta\right|^{4}.\label{eq:LagrangianGL}
\end{equation}
Physically, the expansion coefficient $Z(\lambda,h)$ presents the
(inverse) mass of Cooper pairs. A superconducting transition takes
place when the global minimum of the action $\mathcal{S}_{\textrm{GL}}=\int d\mathbf{x}d\tau\mathcal{L}[\Delta(\mathbf{x},\tau)]$
is located at a nonzero pairing field $\left\langle \Delta(\mathbf{x},\tau)\right\rangle \neq0$,
which describes a condensate of pairs. When $Z(\lambda,h)$ is negative,
Cooper pairs of the lowest energy would have a finite momentum $q_{0}=\sqrt{-Z/(2D)}$,
in order to minimize the combined gradient (kinetic energy) term,
$Zq^{2}+Dq^{4}$ \citep{Pisarski2019}. As a result, the parameter
$r$ effectively renormalizes to $\bar{r}=r-Z^{2}/(4D)$ and a second-order
transition occurs at a critical altermagnetic coupling $\lambda_{c}$
determined by $\bar{r}(h,\lambda_{c})=0$ for a fixed magnetic field
$h$. This mechanism underlies the onset of the altermagnetism-induced
FFLO state illustrated in Fig. \ref{fig1_instability}(a). Conversely,
when $Z(\lambda,h)$ is positive, the pairs have the lowest energy
at zero momentum $q_{0}=0$. In this case, a uniform superconducting
state emerges as the pairing field $\Delta$ becomes constant, with
the transition occurring at $r(h,\lambda_{c})=0$. This behavior corresponds
to the BCS-like state observed at $\lambda=0.70$ in Fig. \ref{fig1_instability}(b).
However, an interesting question arises: why does a nonzero $q_{\max}$
appear when $\lambda$ is increased or decreased from $0.70$, suggesting
negative $Z$? A simple explanation is that the coefficient $Z(h,\lambda)$
may change sign as a function of $\lambda$, thereby altering the
momentum structure of the lowest-energy Cooper pairs.

\begin{figure}
\begin{centering}
\includegraphics[width=0.5\textwidth]{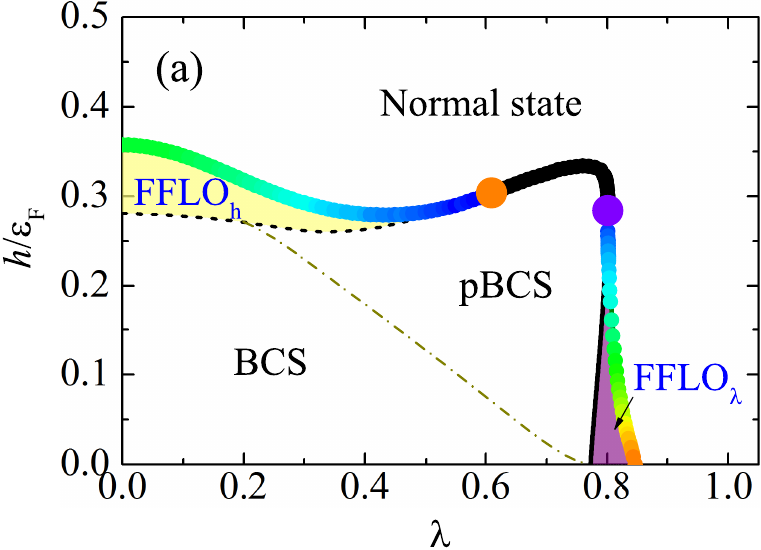}
\par\end{centering}
\begin{centering}
\includegraphics[width=0.5\textwidth]{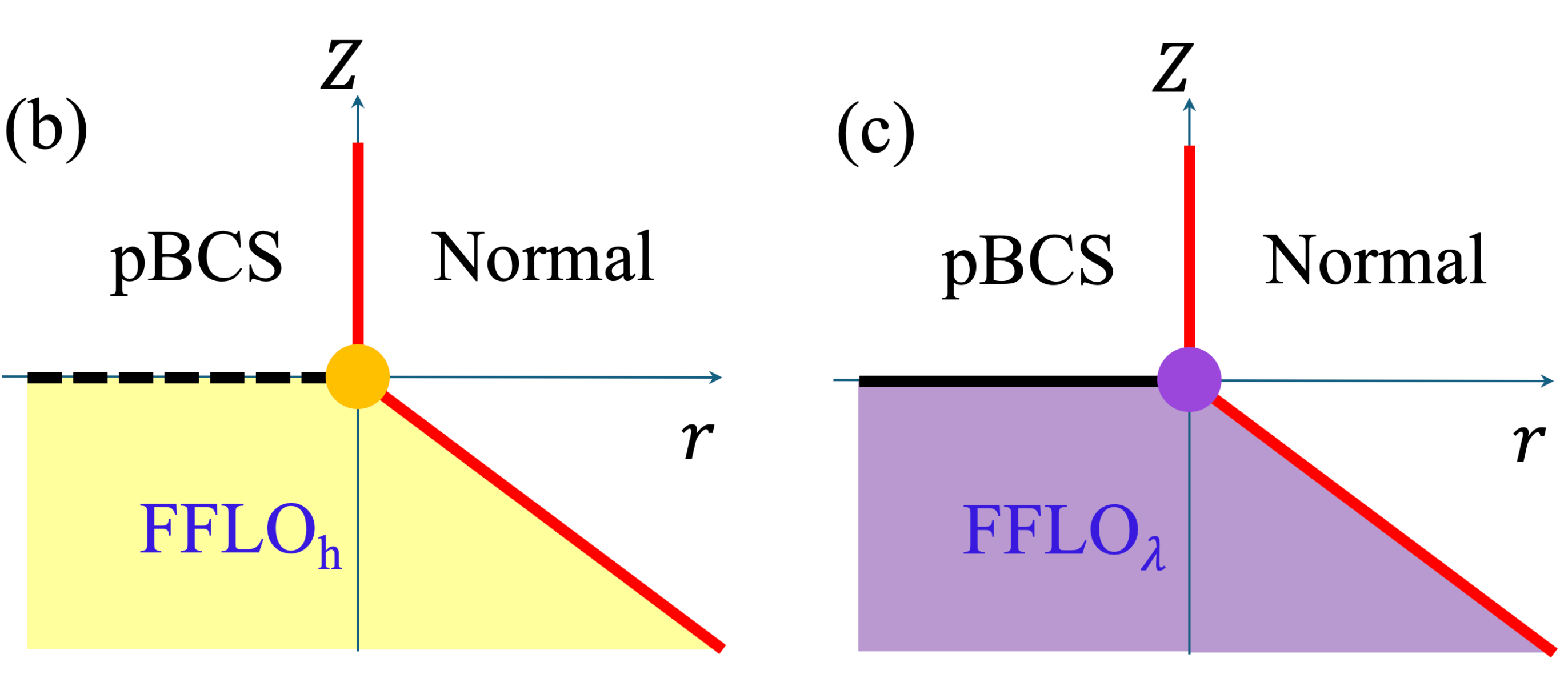}
\par\end{centering}
\caption{\label{fig3_phasediagram} (a) The phase diagram determined by the
mean-field calculations, which further reveal the existence of the
standard BCS phase, a polarized BCS phase and two FFLO states inside
the superconducting phase. This leads to two quantum Lifshitz points,
as highlighted by the orange and purple dots, respectively. (b) and
(c) A sketch of phase diagrams near the quantum Lifshitz points, constructed
by using the phenomenological parameters $Z$ and $r$ following the
effective action. The solid lines and dashed lines in the phase diagrams
correspond to the second-order and first-order phase transitions,
respectively.}
\end{figure}

This leads to an intriguing possibility: by appropriately tuning $h$
and $\lambda$, both coefficients $r(h,\lambda)$ and $Z(h,\lambda)$
might simultaneously vanish at certain tri-critical Lifshitz points
($h_{\textrm{LP}}$, $\lambda_{\textrm{LP}}$), at which two ordered
phases (i.e., the FFLO state and BCS-like state) merge with a disordered,
normal phase. To confirm such an anticipation, we present contour
plots in Fig. \ref{fig2_invg0_qmax}(a) and Fig. \ref{fig2_invg0_qmax}(b),
which show the maximum of the inverse vertex function $\Gamma^{-1}(q_{\max},\omega=0)$
and the corresponding pair momentum $q_{\max}$ in the $\lambda$-$h$
plane, respectively. In the context of our low-energy GL theory, the
former serves as a proxy for the coefficient $-r$ (or $-\bar{r}$),
while the latter can be interpreted as $q_{0}$. Remarkably, Fig.
\ref{fig2_invg0_qmax}(b) reveals a dark stripe, indicating a region
where the pairing momentum is strictly zero. Fig. \ref{fig2_invg0_qmax}(c)
outlines the boundaries of this stripe using two dot-dashed lines
and includes an instability line determined by the Thouless criterion.
These dot-dashed lines and the instability line correspond to the
conditions $Z=0$ and $r=0$ (or $\bar{r}=0$), respectively. Their
intersections mark the anticipated tri-critical Lifshitz points. In
Fig. \ref{fig3_phasediagram}, we highlight the two resulting Lifshitz
points with orange and purple dots.

It is worth noting that the two predicted Lifshitz points are quantum
critical points \citep{Sachdev1999}, as our analysis is conducted
at essentially zero temperature. In contrast, most previous studies
focus on classical Lifshitz points, where tuning the temperature is
typically required to drive the Cooper-pair mass $Z$ to zero or negative
values ($Z\leq0$) \citep{Gubbels2009}. In such cases, the transition
to an ordered phase is governed by thermal fluctuations. Even when
a quantum Lifshitz point emerges accidentally at zero temperature,
the two ordered phases are generally separated by a finite-temperature
quantum critical regime \citep{Ramazashvili1999}. In our case, however,
the predicted quantum Lifshitz points are driven purely by quantum
fluctuations, and the two superconducting ordered states are directly
connected in the phase diagram. This makes the altermagnetic metal
a particular promising platform for probing the rich quantum nature
of Lifshitz points. For instance, the dynamic exponent $z=4$ becomes
large precisely at the quantum Lifshitz point, rendering the associated
critical exponents very unusual. A detailed analysis of these critical
exponents, within the framework of the Hertz-Millis theory \citep{Hertz1976,Millis1993}
will be presented in future work.

\textit{Two types of quantum Lifshitz points}. -- Thus far, our analysis
has been limited to the normal state, as we assumed the smallness
of the pairing field in deriving the effective action $\mathcal{S}_{\textrm{eff}}$
in Eq. (\ref{eq:Seff}). To explore the superconducting phases, we
have also applied the mean-field theory to calculate the nonzero order
parameter $\Delta(\mathbf{x})$, as detailed elsewhere \citep{Hu2025}.
The resulting phase diagram is presented in Fig. \ref{fig3_phasediagram}(a),
which clearly distinguishes between different FFLO states induced
by either the external magnetic field or the intrinsic altermagnetism.
These two mechanisms of FFLO formation compete, ultimately giving
rise to the dark stripe of zero pairing momentum identified in Fig.
\ref{fig2_invg0_qmax}(c). Interestingly, beyond the superfluid transition,
the BCS-like superconducting phase manifests as a novel polarized
BCS state characterized by an imbalance in spin population \citep{Hu2025}.

The two predicted quantum Lifshitz points are distinct, as the nature
of the phase transitions between the polarized BCS state and the two
FFLO states differs significantly. For the field-induced FFLO state
(i.e., FFLO$_{\textrm{h}}$), the transition to the polarized BCS
state is first-order. In contrast, the transition from the altermagnetism-driven
FFLO state (i.e., FFLO$_{\lambda}$) to the polarized BCS state is
continuous. We tentatively attribute the discontinuous transition
near the magnetic-field-induced quantum Lifshitz point (indicated
by the orange dot) to the sign change in the quartic coefficient $u$,
as well as the influence of high-order terms in $\left|\Delta\right|^{2}$
in the GL Lagrangian when extended to the superfluid phase.

\begin{figure}
\begin{centering}
\includegraphics[width=0.25\textwidth]{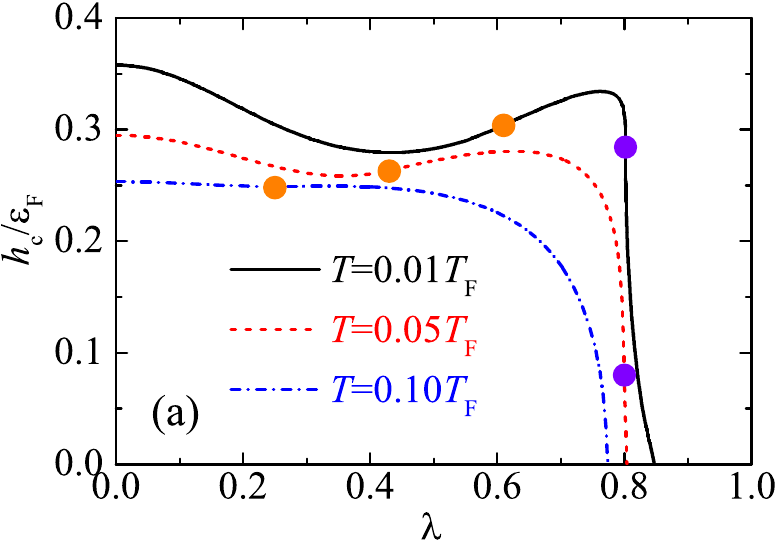}\includegraphics[width=0.25\textwidth]{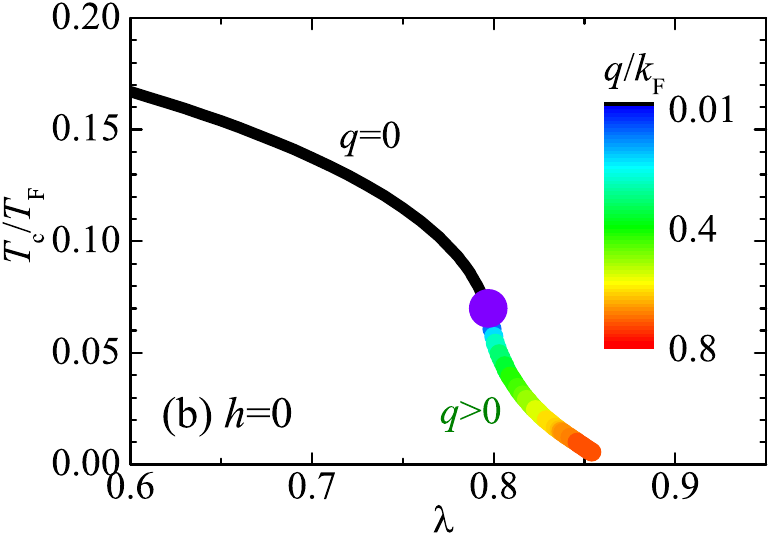}
\par\end{centering}
\caption{\label{fig4_temperature} (a) The instability lines at three temperatures
$T/T_{F}=0.01$, $0.05$ and $0.10$, with the Lifshitz points highlighted
by dots. (b) The critical temperature $T_{c}$ as a function of the
altermagnetic coupling $\lambda$ at zero magnetic field $h=0$. The
color of the line shows the pairing momentum at the phase transition.
The purple dot indicates the zero-field Lifshitz point at $\lambda\simeq0.8$
and at $T\simeq0.07T_{F}$.}
\end{figure}

\textit{Thermal effect on the Lifshitz points}. -- Finally, we briefly
discuss the impact of thermal fluctuations on the Lifshitz points,
as illustrated in Fig. \ref{fig4_temperature}. As the temperature
increases, the Lifshitz point induced by the magnetic field (or by
altermagnetism) shifts toward regions of weaker altermagnetic coupling
(or lower magnetic field). Notably, the altermagnetism-driven Lifshitz
point is more fragile to temperature changes, and disappears entirely
when temperature exceeds approximately $0.07T_{F}$ (for the binding
energy $\varepsilon_{B}=0.08\varepsilon_{F}$ considered in this work),
as shown in Fig. \ref{fig4_temperature}(b).

\textit{Conclusions.} -- We conclude by discussing the experimental
prospects for realizing the predicted quantum Lifshitz points. It
is important to emphasize that our proposed mechanism for generating
quantum Lifshitz points in altermagnetic metals is both robust and
universal -- insensitive to the specific type of altermagnetism or
the particular partial-wave character of the interaction driving superconductivity
\citep{Note}. Experimentally, $d$-wave altermagnetism has been routinely
observed in a variety of materials, with the coupling strength often
tunable. However, superconductivity coexisting with altermagnetism
has yet to be demonstrated. We expect that such superconductivity
may soon be observed in altermagnetic materials featuring strong electron-phonon
coupling, which naturally gives rise to a phonon-mediated $s$-wave
attractive interaction. Alternatively, $d$-wave superconductivity
could emerge in systems with strong spin fluctuations \citep{Scalapino1986},
particularly in materials with significant on-site electronic repulsion.
\begin{acknowledgments}
\textit{Acknowledgments}. -- This research was supported by the Australian
Research Council's (ARC) Discovery Program, Grants Nos. DP240101590
(H.H.) and DP240100248 (X.-J.L.).
\end{acknowledgments}


\begin{thebibliography}{99}
\bibitem{Sachdev1999}S. Sachdev, \textit{Quantum Phase Transitions}
(Cambridge University Press, Cambridge, 1999).

\bibitem{Sachdev2023}S. Sachdev, \textit{Quantum Phases of Matter
}(Cambridge University Press, New York, 2023).

\bibitem{Hornreich1975}R. M. Hornreich M. Luban, and S. Shtrikman
, Critical Behavior at the Onset of $\mathbf{\overrightarrow{k}}$-Space
Instability on the $\lambda$ Line, Phys. Rev. Lett. \textbf{35},
1678 (1975).

\bibitem{Sak1978}J. Sak and G. S. Grest, Critical exponents for the
Lifshitz point: $\varepsilon$-expansion, Phys. Rev. B \textbf{17},
3602 (1978).

\bibitem{Mergulhao1999}C. Mergulhão, Jr. and C. E. I. Carneiro, Field-theoretic
calculation of critical exponents for the Lifshitz point, Phys. Rev.
B \textbf{59}, 13954 (1999).

\bibitem{Diehl2000}H. W. Diehl and M. Shpot, Critical behavior at
$m$-axial Lifshitz points: Field-theory analysis and $\varepsilon$-expansion
results, Phys. Rev. B \textbf{62}, 12338 (2000).

\bibitem{Pisarski2019}R. D. Pisarski, V. V. Skokov, and A. M. Tsvelik,
A pedagogical introduction to the Lifshitz regime, Universe \textbf{5},
48 (2019).

\bibitem{Madhogaria2021}R. P. Madhogaria, C.-M. Hung, B. Muchharla,
A. T. Duong, R. Das, P. T. Huy, S. Cho, S. Witanachchi, H. Srikanth,
and M.-H. Phan, Strain-modulated helimagnetism and emergent magnetic
phase diagrams in highly crystalline MnP nanorod films, Phys. Rev.
B \textbf{103}, 184423 (2021).

\bibitem{Fredrickson1997}G. H. Fredrickson and F. S. Bates, Design
of Bicontinuous Polymeric Microemulsions. J. Polym. Sci. \textbf{35},
2775 (1997).

\bibitem{Jones2012}B. H. Jones and T. P. Lodge, Nanocasting nanoporous
inorganic and organic materials from polymeric bicontinuous microemulsion
templates. Polym. J. \textbf{44}, 131 (2012).

\bibitem{Musevic1982}I. Muševi\v{c}, B. Žekš, and R. Blinc Th. Rasing
and P. Wyder, Phase Diagram of a Ferroelectric Chiral Smectic Liquid
Crystal near the Lifshitz Point, Phys. Rev. Lett. \textbf{48}, 192
(1982).

\bibitem{Gubbels2009}K. B. Gubbels, J. E. Baarsma, and H. T. C. Stoof,
Lifshitz Point in the Phase Diagram of Resonantly Interacting $^{6}$Li\textminus $^{40}$K
Mixtures, Phys. Rev. Lett. \textbf{103}, 195301 (2009).

\bibitem{Casalbuoni2004}R. Casalbuoni and G. Nardulli, Inhomogeneous
superconductivity in condensed matter and QCD, Rev. Mod. Phys. \textbf{76},
263 (2004).

\bibitem{Ramazashvili1999}R. Ramazashvili, Quantum Lifshitz point,
Phys. Rev. B 60, 7314 (1999).

\bibitem{Gunther2007}F. Günther, G. Seibold, and J. Lorenzana, Quantum
Lifshitz Point in the Infinite-Dimensional Hubbard Model, Phys. Rev.
Lett. \textbf{98}, 176404 (2007).

\bibitem{Balents2016}L. Balents and O. A. Starykh, Quantum Lifshitz
Field Theory of a Frustrated Ferromagnet, Phys. Rev. Lett. \textbf{116},
177201 (2016).

\bibitem{Wardh2018}J. Wårdh, B. M. Andersen, and M. Granath, Suppression
of superfluid stiffness near a Lifshitz-point instability to finite-momentum
superconductivity, Phys. Rev. B \textbf{98}, 224501 (2018).

\bibitem{Zhao2020}B. Zhao, J. Takahashi, and A. W. Sandvik, Multicritical
Deconfined Quantum Criticality and Lifshitz Point of a Helical Valence-Bond
Phase, Phys. Rev. Lett. \textbf{125}, 257204 (2020).

\bibitem{Kharkov2020}Y. A. Kharkov, J. Oitmaa, and O. P. Sushkov,
Quantum Lifshitz criticality in a frustrated two-dimensional XY model,
Phys. Rev. B \textbf{101}, 035114 (2020).

\bibitem{Zdybel2020}P. Zdybel and P. Jakubczyk, Quantum Lifshitz
points and fluctuation-induced first-order phase transitions in imbalanced
Fermi mixtures, Phys. Rev. Res. \textbf{2}, 033486 (2020).

\bibitem{Hayami2019}S. Hayami, Y. Yanagi, and H. Kusunose, Momentum-Dependent
Spin Splitting by Collinear Antiferromagnetic Ordering, J. Phys. Soc.
Jpn. \textbf{88}, 123702 (2019).

\bibitem{Hayami2020a}S. Hayami, Y. Yanagi, and H. Kusunose, Spontaneous
antisymmetric spin splitting in noncollinear antiferromagnets without
spin-orbit coupling, Phys. Rev. B \textbf{101}, 220403(R) (2020).

\bibitem{Hayami2020b}S. Hayami, Y. Yanagi, and H. Kusunose, Bottom-up
design of spin-split and reshaped electronic band structures in antiferromagnets
without spin-orbit coupling: Procedure on the basis of augmented multipoles,
Phys. Rev. B \textbf{102}, 144441 (2020).

\bibitem{Smejkal2020}L. Šmejkal, R. González-Hernández, T. Jungwirth,
and J. Sinova, Crystal time-reversal symmetry breaking and spontaneous
Hall effect in collinear antiferromagnets, Sci. Adv. \textbf{6}, eaaz8809
(2020).

\bibitem{Smejkal2022}L. Šmejkal, J. Sinova, and T. Jungwirth, Emerging
Research Landscape of Altermagnetism, Phys. Rev. X \textbf{12}, 040501
(2022).

\bibitem{Mazin2023}I. I. Mazin, Altermagnetism in MnTe: Origin, predicted
manifestations, and routes to detwinning, Phys. Rev. B \textbf{107},
L100418 (2023).

\bibitem{Das2024}P. Das, V. Leeb, J. Knolle, and M. Knap, Realizing
Altermagnetism in Fermi-Hubbard Models with Ultracold Atoms, Phys.
Rev. Lett. \textbf{132}, 263402 (2024).

\bibitem{Hiraishi2024}M. Hiraishi, H. Okabe, A. Koda, R. Kadono,
T. Muroi, D. Hirai, and Z. Hiroi, Nonmagnetic ground state in RuO2
revealed by muon spin rotation, Phys. Rev. Lett. \textbf{132}, 166702
(2024).

\bibitem{Amin2024}O. J. Amin, A. Dal Din, E. Golias, Y. Niu, A. Zakharov,
S. C. Fromage, C. J. B. Fields, S. L. Heywood, R. B. Cousins, F. Maccherozzi,
J. Krempaský, J. H. Dil, D. Kriegner, B. Kiraly, R. P. Campion, A.
W. Rushforth, K. W. Edmonds, S. S. Dhesi, L. Šmejkal, T. Jungwirth,
and P. Wadley, Nanoscale imaging and control of altermagnetism in
MnTe, Nature \textbf{636}, 348 (2024).

\bibitem{McClarty2024}P. A. McClarty and J. G. Rau, Landau Theory
of Altermagnetism, Phys. Rev. Lett. \textbf{132}, 176702 (2024).

\bibitem{Lu2024}B. Lu, K. Maeda, H. Ito, K. Yada, and Y. Tanaka,
\ensuremath{\varphi} Josephson Junction Induced by Altermagnetism,
Phys. Rev. Lett. \textbf{133}, 226002 (2024).

\bibitem{Maeda2024}K. Maeda, B. Lu, K. Yada, and Y. Tanaka, Theory
of Tunneling Spectroscopy in Unconventional $p$-Wave Magnet-Superconductor
Hybrid Structures, J. Phys. Soc. Jpn. \textbf{93}, 114703 (2024).

\bibitem{Fukaya2025a}Y. Fukaya, K. Maeda, K. Yada, J. Cayao, Y. Tanaka,
and B. Lu, Josephson effect and odd-frequency pairing in superconducting
junctions with unconventional magnets, Phys. Rev. B \textbf{111},
064502 (2025).

\bibitem{Fukaya2025b}Y. Fukaya, B. Lu, K. Yada, Y. Tanaka, and J.
Cayao, Superconducting phenomena in systems with unconventional magnets,
arXiv:2502.15400 (2025).

\bibitem{Lin2025}H.-J. Lin, S.-B. Zhang, H.-Z. Lu, and X.\LyXThinSpace C.
Xie, Coulomb Drag in Altermagnets, Phys. Rev. Lett. \textbf{134},
136301 (2025).

\bibitem{Chakraborty2024}D. Chakraborty and A. M. Black-Schaffer,
Zero-field finite-momentum and field-induced superconductivity in
altermagnets, Phys. Rev. B \textbf{110}, L060508 (2024).

\bibitem{Hong2025}S. Hong, M. J. Park, and K.-M. Kim, Unconventional
$p$-wave and finite-momentum superconductivity induced by altermagnetism
through the formation of Bogoliubov Fermi surface, Phys. Rev. B \textbf{111},
054501 (2025).

\bibitem{Fulde1964}P. Fulde and R. A. Ferrell, Superconductivity
in a Strong Spin-Exchange Field, Phys. Rev. \textbf{135}, A550 (1964).

\bibitem{Larkin1964}A. I. Larkin and Yu. N. Ovchinnikov, Nonuniform
state of superconductors, Zh. Eksp. Teor. Fiz. \textbf{47}, 1136 (1964)
{[}Sov. Phys. JETP \textbf{20}, 762 (1965){]}.

\bibitem{Ohashi2002}Y. Ohashi, On the Fulde--Ferrell State in Spatially
Isotropic Superconductors, J. Phys. Soc. Jpn. \textbf{71}, 2625 (2002).

\bibitem{Uji2006}S. Uji, T. Terashima, M. Nishimura, Y. Takahide,
T. Konoike, K. Enomoto, H. Cui, H. Kobayashi, A. Kobayashi, H. Tanaka,
M. Tokumoto, E. S. Choi, T. Tokumoto, D. Graf, and J. S. Brooks, Vortex
Dynamics and the Fulde-Ferrell-Larkin-Ovchinnikov State in a Magnetic-Field-Induced
Organic Superconductor, Phys. Rev. Lett. \textbf{97}, 157001 (2006).

\bibitem{Hu2006}H. Hu and X.-J. Liu, Mean-field phase diagrams of
imbalanced Fermi gases near a Feshbach resonance, Phys. Rev. A \textbf{73},
051603(R) (2006).

\bibitem{Hu2007}H. Hu, X.-J. Liu, and P. D. Drummond, Phase Diagram
of a Strongly Interacting Polarized Fermi Gas in One Dimension, Phys.
Rev. Lett. \textbf{98}, 070403 (2007).

\bibitem{Sheehy2015}D. E. Sheehy, Fulde-Ferrell-Larkin-Ovchinnikov
state of two-dimensional imbalanced Fermi gases, Phys. Rev. A \textbf{92},
053631 (2015).

\bibitem{He2015}L. He, H. Lü, G. Cao, H. Hu, and X.-J. Liu, Quantum
fluctuations in the BCS-BEC crossover of two-dimensional Fermi gases,
Phys. Rev. A \textbf{92}, 023620 (2015).

\bibitem{SadeMelo1993}C. A. R. Sá de Melo, M. Randeria, and J. R.
Engelbrecht, Crossover from BCS to Bose superconductivity: Transition
temperature and time-dependent Ginzburg-Landau theory, Phys. Rev.
Lett. \textbf{71}, 3202 (1993).

\bibitem{Liu2006}X.-J. Liu and H. Hu, BCS-BEC crossover in an asymmetric
two-component Fermi gas, Europhys. Lett. \textbf{75}, 364 (2006).

\bibitem{Hertz1976}J. A. Hertz, Quantum critical phenomena, Phys.
Rev. B \textbf{14}, 1176 (1976).

\bibitem{Millis1993}A. J. Millis, Effect of a nonzero temperature
on quantum critical points in itinerant fermion systems, Phys. Rev.
B \textbf{48}, 7183 (1993).

\bibitem{Hu2025}H. Hu, Z. Liu, and X.-J. Liu, Unconventional superconductivity
of an altermagnetic metal: Polarized BCS and inhomogeneous Fulde-Ferrell-Larkin-Ovchinnikov
states, arXiv:2505.10196v1 (2025).

\bibitem{Note}We have investigated the presence of quantum Lifshitz
points in systems with $d_{xy}$-wave and $d_{x^{2}-y^{2}}$-wave
altermagnetism, considering both $s$-wave and $d$-wave attractive
interactions. These findings indicate that the spin-splitting in the
dispersion relation induced by altermagnetism plays the dominant role.

\bibitem{Scalapino1986}D. J. Scalapino, E. Loh, Jr., and J. E. Hirsch,
$d$-wave pairing near a spin-density-wave instability, Phys. Rev.
B \textbf{34}, 8190(R) (1986).
\end{thebibliography}
\end{document}